\documentclass[%
 reprint,
 amsmath,amssymb,
 aps,
pra,
]{revtex4-1}

\usepackage[noend]{algpseudocode}

\usepackage{graphicx}% Include figure files
\usepackage{dcolumn}% Align table columns on decimal point
\usepackage{bm}% bold math
\usepackage{enumitem}
\usepackage{placeins}
\usepackage{xcolor}
\usepackage{hyperref}% add hypertext capabilities
\usepackage{mathtools}

\DeclarePairedDelimiter\floor{\lfloor}{\rfloor}

\newcommand{\diff}[2]{\frac{\partial #1}{\partial #2}}

\begin{document}

%\preprint{APS/123-QED}

\title{Approaching the Quantum Speed Limit with Global-Local Optimization}% Force line breaks with \\

\author{J. J. W. H. S\o rensen}
\author{M. O. Aranburu}
\author{T. Heinzel}
\author{J. F. Sherson}
\affiliation{Aarhus University}
\email{sherson@phys.au.dk}
\date{\today}% It is always \today, today,
             %  but any date may be explicitly specified
\begin{abstract}

We propose a Global-Local optimization algorithm for quantum control that combines standard local search methodologies with evolutionary algorithms. This allows us to find faster solutions to a set of problems relating to ultracold control of Bose-Einstein condensates.
\end{abstract}

\maketitle

%\tableofcontents

\section{Introduction}
Quantum engineering aims to control and steer quantum dynamics in order to realize specific quantum states or operations. This has numerous applications in e.g. femtosecond lasers \cite{assion1998control,meshulach1998coherent}, quantum gate synthesis \cite{schulte2005optimal} and quantum many-body systems \cite{doria2011optimal}. These applications often rEquire tailored control pulses that precisely manipulate the quantum dynamics. Due to experimental limitations such as decoherence, the control pulses must typically be as fast as possible \cite{van2016optimal}. The search for the most time-optimal control or Quantum Speed Limit (QSL) has attracted much attention in the literature \cite{levitin2009fundamental,taddei2013quantum,caneva2009optimal,deffner2017quantum}. 

Quantum Optimal Control (QOC) is a tool that finds control pulses by reformulating the control problem as an optimization problem \cite{werschnik2007quantum}. If there are few constraints and full controllability, then these optimization problems are benign in the sense that all local maxima are also global maxima \cite{russell2016quantum,rabitz2004quantum}. These optimization problems can be solved with local "hill climber" type algorithms, since they converge towards to a local and thereby global maxima. However, when we seek fast solutions one must introduce a low bound on the total process duration (\textit{T}). This constraint removes the benign properties of the control problem and local algorithms are no longer guaranteed to find global maxima \cite{zhdanov2015role,bukov2017machine}.

\begin{figure}[ht]
\includegraphics[width=0.80\columnwidth]{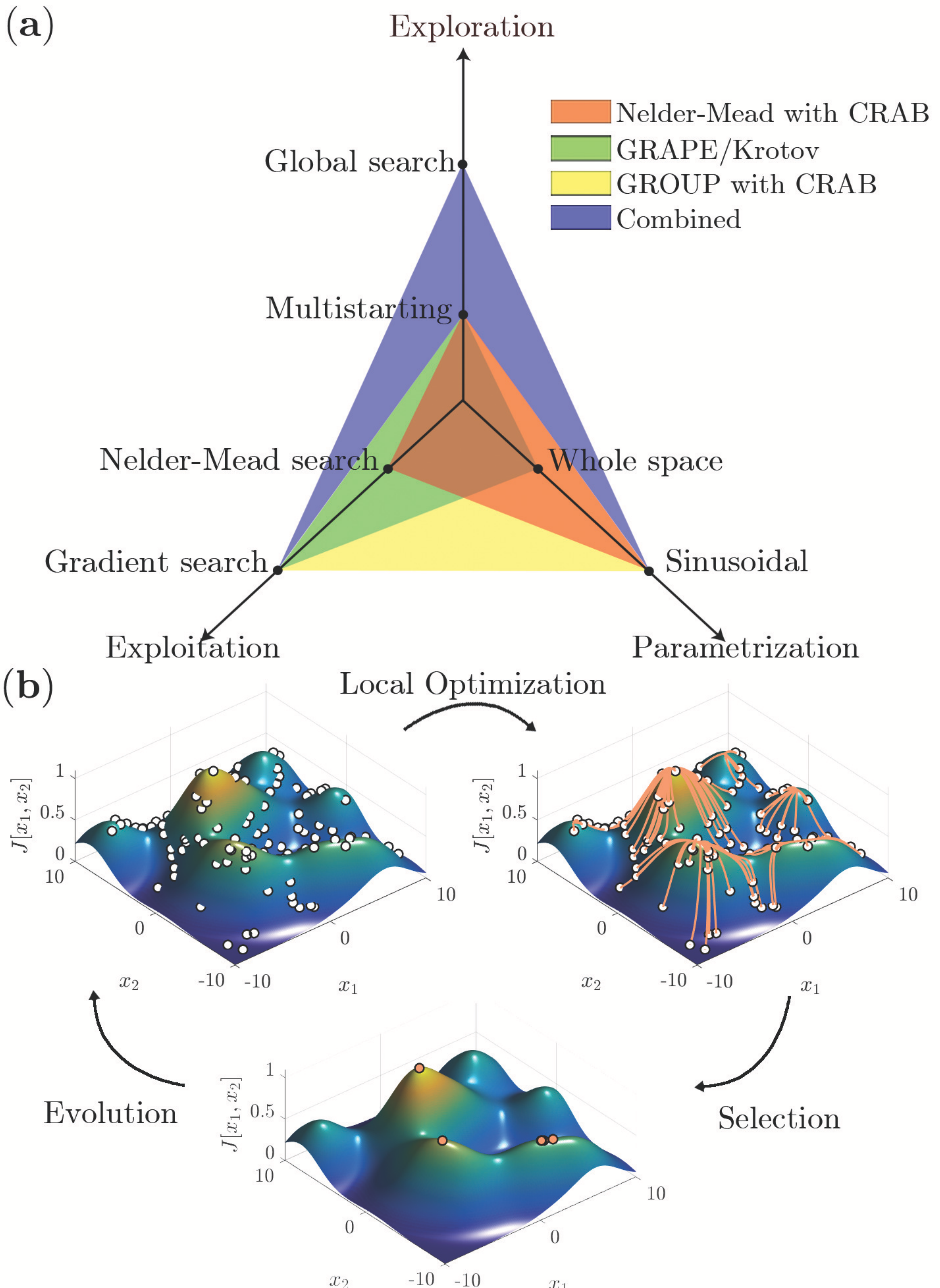}
\centering
\caption{(\textbf{a}) The combined search has three main components namely global and local search methods and the parametrization of the control. It is shown how other standard methods are placed within this categorization. (\textbf{b}) The main steps in the Global-Local algorithm. Here the problem has been pictured as a maximization for ease of visualization. }
\label{fig:cartoon}
\end{figure}

These considerations show that finding the precise location of the QSL can be a difficult optimization problem. Solving such problems require consideration of three main aspects: exploration, exploitation and problem parametrization - see Fig. \ref{fig:cartoon}\textbf{a}. Exploration is searching for new candidate solutions in sparsely probed parts of the control space, whereas exploitation is intense analysis of a small portion of the control space enhancing the best solution \cite{neri2012memetic}. Finally, the control space is often high dimensional. This high dimension can be reduced by a proper problem parametrization, which eases the search for optimal solutions \cite{localPaper}.

Traditionally, in QOC emphasis is placed on purely exploitative local search algorithms like \textsc{grape} and Nelder-Mead with \textsc{crab} that respectively find solutions using derivative-information or search in a reduced control basis \cite{khaneja2005optimal,caneva2011chopped}. Recently, we introduced the \textsc{group} optimization algorithm that does a gradient descent in a reduced basis and thereby combines the advantages of \textsc{crab} and \textsc{grape} \cite{localPaper}. For these local algorithms, basic exploration is typically added on top of the local search using multistarting \cite{brouzos2015quantum,doria2011optimal}. Modern algorithms from computer science like Differential Evolution (\textsc{de}) dynamically adjust the balance between exploration and exploitation to increase performance. However, they lack domain specific features such as analytic gradient expressions built into the standard algorithms.

Here we propose a hybrid Global-Local algorithm that combines \textsc{de} with local algorithms. This gives a better balance between exploration and exploitation while retaining the domain specific features. We parametrize the control with sinusoidal functions as in the \textsc{crab}-method. As shown in Fig. \ref{fig:cartoon}\textbf{a} this algorithm better balances all three main aspects. We apply this method to control of Bose-Einstein condensates (BECs) in Condensate Splitting (CS) and Condensate Driving (CD). Here we observe improvements in the estimate of the QSL. Below the quantum speed limit there is a conjectured universal $\sin^2$-behavior of the fidelity as a function of duration ($F(T)$) \cite{caneva2009optimal,gajdacz2015time}. In recent work in Refs. \cite{sorensen2016exploring,gajdacz2015time} we cast doubt on the generality of this conjecture and the $F(T)$-curves presented in this work further strengthen this doubt.

This paper is organized as follows: In section \ref{sec:CostFunLocalOpt} we briefly present the local optimization used in the combined Global-Local algorithm. In section \ref{sec:GloLloOpt} we present our proposed combined Global-Local algorithm and we apply it in two controls problems in section \ref{sec:Results}. In section \ref{sec:Learning} we compare our proposed algorithm with conventional multistarting algorithms.

\begin{figure*}[t]
\begin{minipage}[t]{.48\textwidth}
  \includegraphics[width=\textwidth]{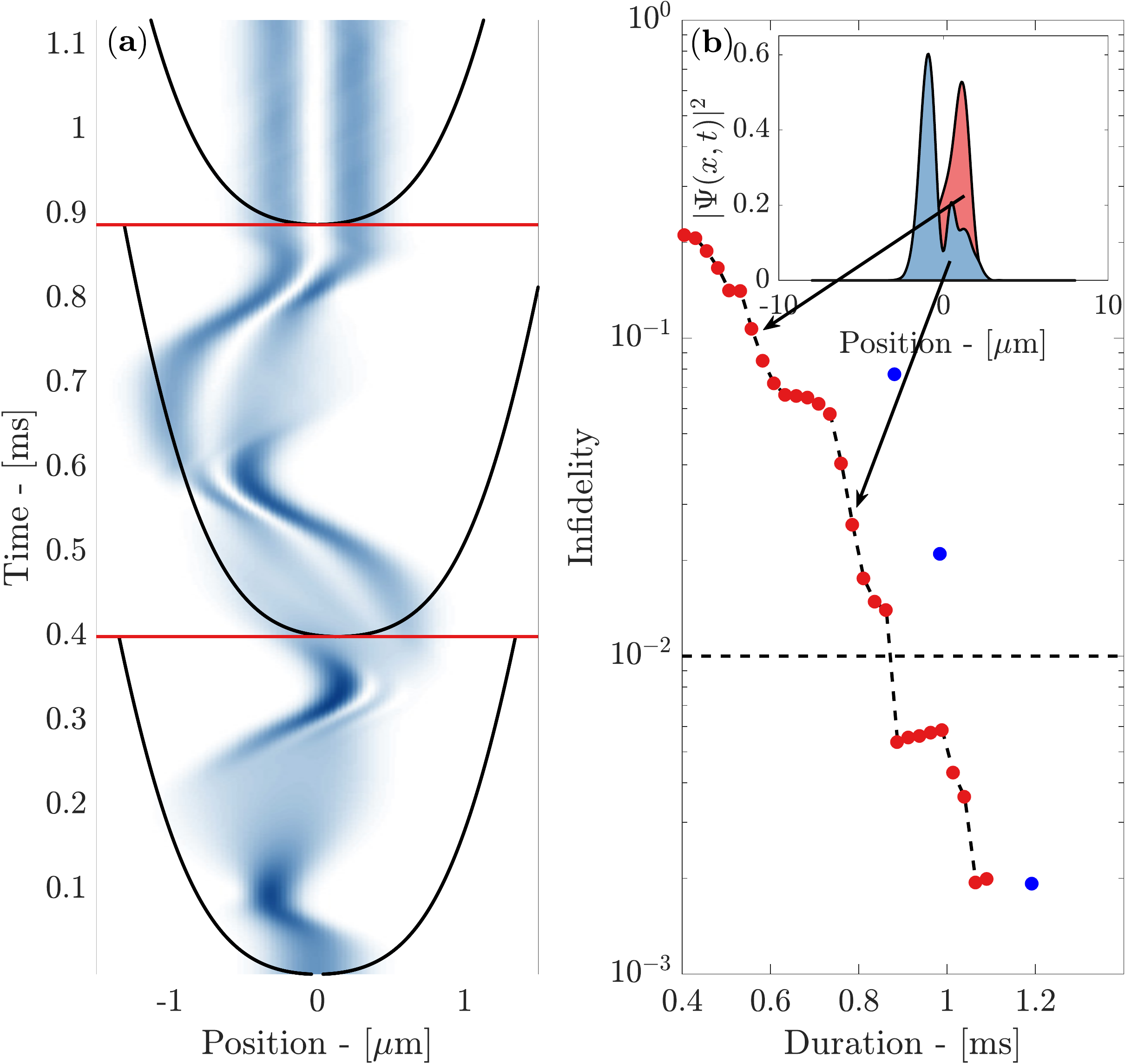}
  \caption{Condensate Driving (CD). (\textbf{a}) The density of the condensate ($|\psi(x,t)|^2$) when propagated along the best control at a duration of $T=0.89 \text{ms}$. Above the top line, the potential is held constant and the state is seen to be stationary. The associated potentials are drawn on top for $t=0$,$t=0.4T$ and $t=T$ where \textit{T} is the process duration. (\textbf{b}) The best infidelity at different durations from the Global-Local algorithm shown using red dots. The blue dots show some of results from the $F(T)$-curve reported in Ref. \cite{van2016optimal}. An insert shows the density of the wavefunction at two different durations.}
  \label{fig:antonioCondensate}
\end{minipage}
\quad
\begin{minipage}[t]{.48\textwidth}
  \includegraphics[width=\textwidth]{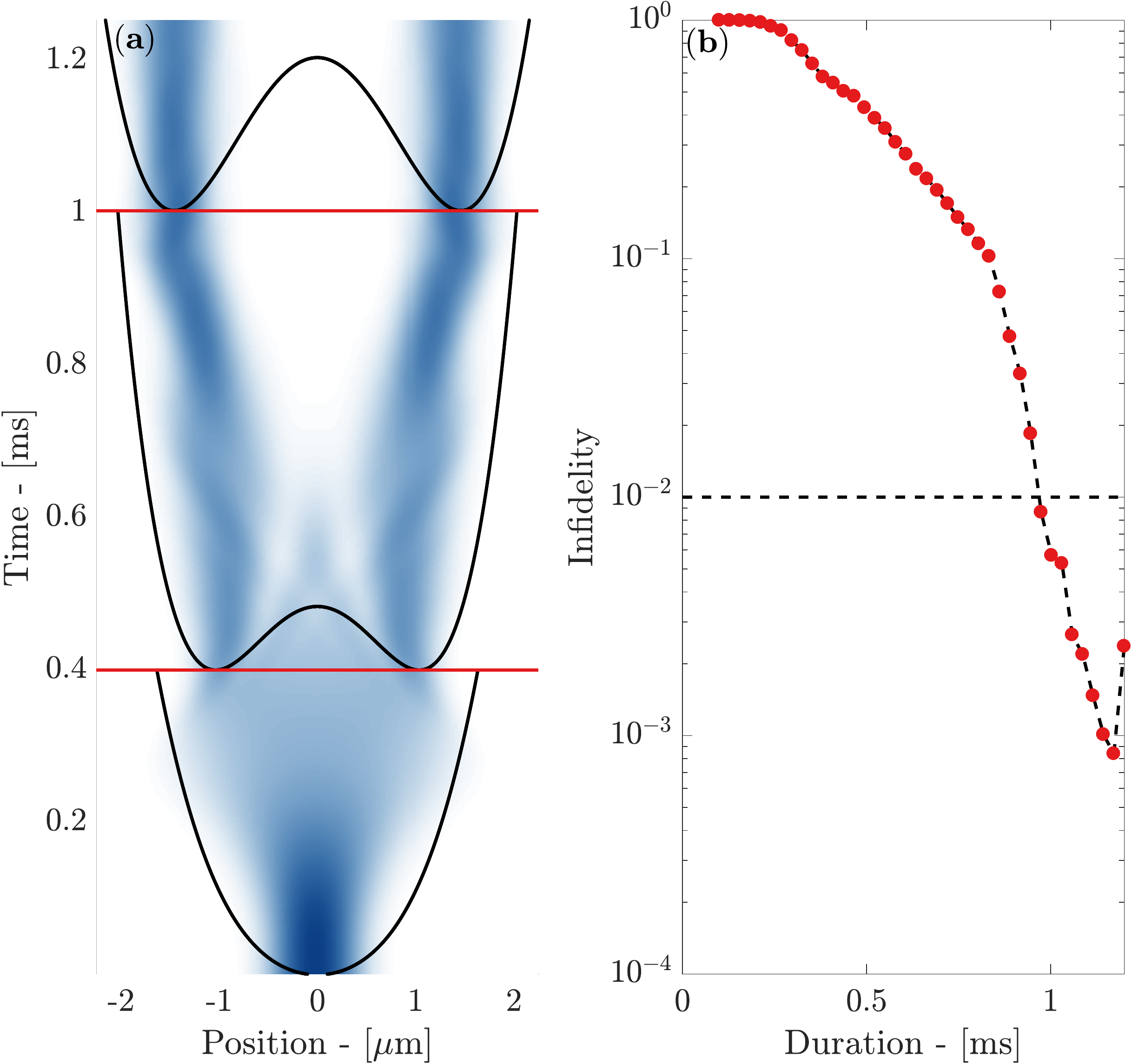}
  \caption{Condensate Splitting (CS). (\textbf{a}) this shows the condensate density ($|\psi(x,t)|^2$) along the best control at a duration of $T=1.0 \text{ms}$. The potential is held constant above the top line, which shows that the state is in an eigenstate. The potential is shown for durations $t=0$,$t=0.4T$ and $t=T$. (\textbf{b}) Infidelity shown for different durations obtained using the Global-Local algorithm displayed with red dots.}
\label{fig:lesanovskyCondensate}
\end{minipage}
\end{figure*}

\section{Cost Functional and Local Optimization} \label{sec:CostFunLocalOpt}
Before we discuss the in-depth structure of the Global-Local algorithm we briefly present the BEC control problems. The dynamics of a BEC can be described in a mean-field by the Gross-Pitaevskii Equation (GPE)
\begin{align}
    i\diff{\psi}{t}&=-\frac{1}{2m}\diff{^2\psi}{t^2}+V(x,u)\psi+\beta |	\psi|^2 \psi \\
    					&=\bigl(\hat{H}+\beta |\psi|^2\bigr) \psi, \label{GPE}
\end{align}
where $\hbar=1$, $\beta$ is the non-linear self-interaction and $\hat{H}$ is the Hamiltonian. Here we assume the system is one-dimensional since the two other spatial directions can be frozen out \cite{van2016optimal}. The potential depends on the control \textit{u}, and the specific expressions for the different potentials are presented in section \ref{sec:Results}. The objective is to transfer the initial state $\psi_0$ into the target state $\psi_t$, which are both eigenstates of the GPE for $u(t=0)$ and $u(t=T)$. This is a so-called state-to-state problem \cite{schirmer2011efficient}.

In QOC such control problems are expressed as a minimization of the cost functional
\begin{equation}
\hat{J}(u)=\frac{1}{2}(1-F)+\frac{\gamma}{2}\int_0^T \dot{u}(t)^2 \text{d}t, \label{costFun}
\end{equation}
where $F=|\langle \psi_T|\psi(T)\rangle|^2$ is the fidelity and $1-F$ is the infidelity, which characterizes the deviation of the final state $\psi(T)$ from the target state $\psi_t$. Here $\hat{J}(u)$ is the so-called reduced cost functional so the states $\psi(t)$ depend implicitly on $u(t)$ through the GPE \cite{localPaper}. The final state $\psi(T)$ is found by solving the GPE with the potential defined by $u(t)$. The second term enforces regularization, which smooths out the control. This accounts for the fact that arbitrarily fast changes in the control cannot be realized experimentally. Typically, a weight $\gamma$ of about $1\cdot 10^{-6}$ is sufficient to achieve an acceptable regularization.

Analytically calculating the minimum of Eq. (\ref{costFun}) is typically not feasible. Instead the standard approach in QOC is to use local iterative optimization algorithms in order to find solutions for Eq. (\ref{costFun}) \cite{hohenester2007optimal,mennemann2015optimal}.

In order to perform the optimization numerically the control $u(t)$ is typically discretized in steps of $\Delta t$, where $\Delta t$ is set by the required accuracy when numerically solving the GPE. This gives an effective dimension for the simulation at $N= \floor{T/\Delta t} $. Often, this dimension is larger than the required dimension for the control problem, since the optimal controls can be expressed in a basis with a smaller dimension \cite{lloyd2014information,localPaper,caneva2011chopped}. This motivates expanding the control in a chopped basis, 
\begin{equation}
u(t) = u_0(t) + S(t) \sum_{n=1}^M c_n f_n(t), \label{crabExpansion}
\end{equation}
where \textit{M} is the size of the basis. As an example we here use $f_n=\sin((n+r_n)\pi t)/T))$ are the basis function with $-0.5\leq r_n \leq 0.5$ being a randomly selected frequency shifts. Here $0\leq S(t) \leq 1$ is a shape function that ensures $u(t=0)=u_0$ and $u(t=T)=u_T$ so $S(0)=S(T)=0$. Within QOC using a random chopped basis was originally introduced in the \textsc{crab} methodology \cite{caneva2011chopped,doria2011optimal}. With this expansion the optimization is performed over the expansion coefficients, so the optimization is for $\hat{J}(\mathbf{c})$ where $\mathbf{c}=(c_1,c_2,...,c_M)$ \cite{localPaper}. Optimization is typically done in this basis using the derivative-free method Nelder-Mead \cite{caneva2011chopped,doria2011optimal}. In Ref. \cite{localPaper} we introduced Gradient Optimization Using Parametrization (\textsc{group}) method that performs a gradient descent within this basis and showed it is competitive with standard QOC algorithms. Here the gradient is calculated using the analytic expression,
\begin{equation}
\diff{\hat{J}(\mathbf{c})}{c_n}=-\int_0^T \biggl(\Re \biggl\langle \chi\biggl|\diff{\hat{H}}{u}\biggr|\psi\biggr\rangle + \gamma \ddot{u} \biggr)S(t)f_n(t)\text{d}t, \label{groupGRAD}
\end{equation}
where $\chi$ is a Lagrange multiplier, which satisfies the equation of motion
\begin{equation}
i  \dot{\chi} = \bigl(\hat{H}+2\beta|\psi|^2\bigr)\chi+ \beta \psi^2\chi^*.
\end{equation}
When the gradient is calculated using Eq. (\ref{groupGRAD}) the control can be iteratively updated along the gradient $u^{(i+1)}=u^{(i)}-\alpha^{(i)} \nabla \hat{J}\bigl(u^{(i)}\bigr)$ with $i=0,1,2,...$ where \textit{i} is the iteration index \cite{khaneja2005optimal,jager2014optimal,mennemann2015optimal}. An appropriate value for $\alpha^{(i)}$ is found using a step-size algorithm. Instead of searching along the negative gradient we use a quasi-Newton method \cite{localPaper}. An in-depth discussion of \textsc{group} is presented in Ref. \cite{localPaper}. 

Although this method is competitive with standard methods in QOC, it is still a linesearch algorithm so it cannot escape local optima with $F<1$ \cite{localPaper}. In order to add such a capability we combine \textsc{group} with a global optimization algorithm.

\section{Global-Local Optimization} \label{sec:GloLloOpt}
If a local optimization algorithm like \textsc{group} converges to a local suboptimum (a control with $F<1$) then the result may be improved by simply optimizing another initial control. This straightforward way of exploring the optimization landscape is known as multistarting \cite{ugray2007scatter}. Multistarting has a constant probability of success given by how likely it is to randomly select a control that optimizes to $F=1$. Close to the QSL there may be only few global maxima so this probability can be very low \cite{zahedinejad2015high,zahedinejad2014evolutionary}. 

An alternative to multistarting is using a global optimization algorithm like Particle Swarm Optimization, Simulated annealing, Differential Evolution (\textsc{de}) and Covariance Matrix Adaptation Evolutionary Strategy (CMA-ES) \cite{kennedy2011particle,kirkpatrick1983optimization,das2011differential,hansen2003reducing}. However, these are domain general algorithms and they do not have access to domain specific features like the analytic gradients and good parametrization used in \textsc{group}, which are important for finding high fidelity solutions. 

In order to combine these two approaches we propose a combined Global-Local algorithm. Here the global algorithm is a replacement for the multistarting strategy. Based on the performance of past solutions the global algorithm proposes new seeds for the local optimization algorithm. In principle, this type of combination could be done with any global optimization algorithm, but here we focus on \textsc{de} due to its good performance in quantum optimal control problems and general optimization contests \cite{zahedinejad2015high,zahedinejad2014evolutionary,das2011differential}. The algorithm updates a population of points (members) in the optimization landscape $P=\{\mathbf{x}_1,\mathbf{x}_2,....,\mathbf{x}_N\}$. These points must fully characterize the control in the chopped random basis (Eq. (\ref{crabExpansion})) so they consists of both the expansion coefficients and the random frequencies $\mathbf{x}_n = (c_1,c_2,...,c_M,r_1,r_2,...,r_M)$. The algorithm iterates in three main steps being evolution, local optimization, and selection, which are illustrated graphically in Fig. \ref{fig:cartoon}\textbf{b}. Completing all three steps is one generation. First a new trial population ($P_t$) is formed using the evolution strategy from \textsc{de}. We outline the evolution strategy below. In the next step some of the members in $P_t$ are locally optimized. The probability for being optimized ($p(n)$) is given as a sigmoid and the members with the lowest cost have the highest probability of being optimized. If a member $\mathbf{x}_n$ is selected for optimization then it is replaced by the optimized member in $P_t$. The local optimization is \textsc{group} as outlined in the previous section. The optimization is only performed on the expansion coefficients ($c_n$).  Finally, in the last selection step the $\mathbf{x}_n$ member in $P$ is replaced by the $\mathbf{x}_n$ member in $P_t$ if the trial member has a lower cost. A pseudocode for the algorithm is shown in Fig. \ref{fig:GloLlo}.

Before discussing the results from the Global-Local algorithm we give a brief account of the evolution strategy used in \textsc{de}. \textsc{de} randomly selects two distinct members ($\mathbf{x}_{j_1},\mathbf{x}_{j_2}$) and the current best member ($\mathbf{x}_{j_b}$). From these three members a donor vector is given as $\mathbf{v}_n = \mathbf{x}_{j_b}+\mathcal{F}(\mathbf{x}_{j_1}-\mathbf{x}_{j_2})$ where $\mathcal{F}$ is a scaling factor. From this donor vector the \textit{n}'th member in the trial population is found by replacing \textit{L} consecutive values of $\mathbf{x}_n$ with values from $v_n$. The length of \textit{L} is given by a Possion distribution with mean \textit{Cr} and minimum length of one. The starting point of this replacement is random. In our simulations, $\mathcal{F}$ is linearly decreased from 0.4 to 0.1 over the simulation in order to promote early exploration and later exploitation. We also use $Cr = 0.97$.
\begin{figure}
\noindent\hfil\rule{0.5\textwidth}{.4pt}\hfil
\begin{algorithmic}[1]
\State Initialize population $P=\{\mathbf{x}_1,\mathbf{x}_2,....,\mathbf{x}_N\}$ and variables.
\While{$F\leq F_{conv}$ and iter$\leq$MaxIter}
\State Using \textsc{de} create a trial population ($P_t$) from $P$.
\For{ each number \textit{n} in $P_t$}
\If{rand()$\leq p(n)$}
\State Optimize the coefficients in $\mathbf{x}_i$ using \textsc{group}.
\EndIf
\If{$P_t(n)<P(n)$}
\State $P(n) \leftarrow P_t(n)$
\EndIf
\EndFor
\State iter $\leftarrow$ iter + 1.
\EndWhile
\end{algorithmic}
\vspace{-7 pt}
\noindent\hfil\rule{0.5\textwidth}{.4pt}\hfil
\caption{A pseudo-code for the combined Global-Local algorithm.}
\label{fig:GloLlo}
\end{figure}

\section{Results} \label{sec:Results}
Optimal control of BECs of $^{87}\text{Rb}$ atoms trapped atom-chips has been explored by several authors and realized experimentally \cite{bucker2013vibrational,schumm2005matter,van2016optimal,jager2014optimal,hohenester2007optimal} . The dynamics of the two problems are shown in Fig. \ref{fig:antonioCondensate}\textbf{a} and \ref{fig:lesanovskyCondensate}\textbf{a}. We discuss each of the two control problems separately.

\subsection{Condensate Driving}
In Condensate Driving (CD) a BEC must be transferred from the initial ground state of an anharmonic well into the first excited state. This state can be used as a source for stimulated emission of matter waves \cite{bucker2011twin}. The transfer is completed by shaking the trap. Previously, in CD an $F(T)$-curve demonstrating a conjectured double-$\sin^2$-behavior has been reported \cite{van2016optimal}. The potential is well described by the polynomial,
\begin{equation*}
    V(x,u(t))=p_2\bigl(x-u(t)\bigr)^2+p_4\bigl(x-u(t)\bigr)^4+p_6\bigl(x-u(t)\bigr)^6,
\end{equation*}
where the control $u(t)$ is the trap displacement \cite{van2016optimal}. The coefficients are given by $p_2 = 2\pi \hbar \cdot 310/r_0^2 \text{J}/\text{m}^2$,$p_4=2\pi\hbar \cdot 13.6/r_0^4 \text{J}/\text{m}^4$ and $p_6=-2\pi\hbar \cdot 0.0634/r_0^6 \text{J}/\text{m}^6$ with $r_0=172 \text{nm} $\cite{van2016optimal}. The nonlinear coupling constant is $\beta=2.61 \hbar \, \mu \text{m}\,\text{Hz}$ for 700 atoms, which takes corrections for going from the three-dimensional to the one-dimensional case into account \cite{van2016optimal,gerbier2004quasi}. A further complication arises in this control problem, which is the fact that it is necessary to include the finite bandwidth of the control electronics. This effect causes the control to become convolved into the new control $v(t)$ and the atoms experience the potential $V(x,v(t))$. This correction must also be included in the expression for the gradient (Eq. (\ref{groupGRAD})). The procedure for including this effect into local optimization is discussed in Ref. \cite{localPaper}.

\begin{figure*}
\begin{minipage}[t]{.48\textwidth}
\includegraphics[width=\textwidth]{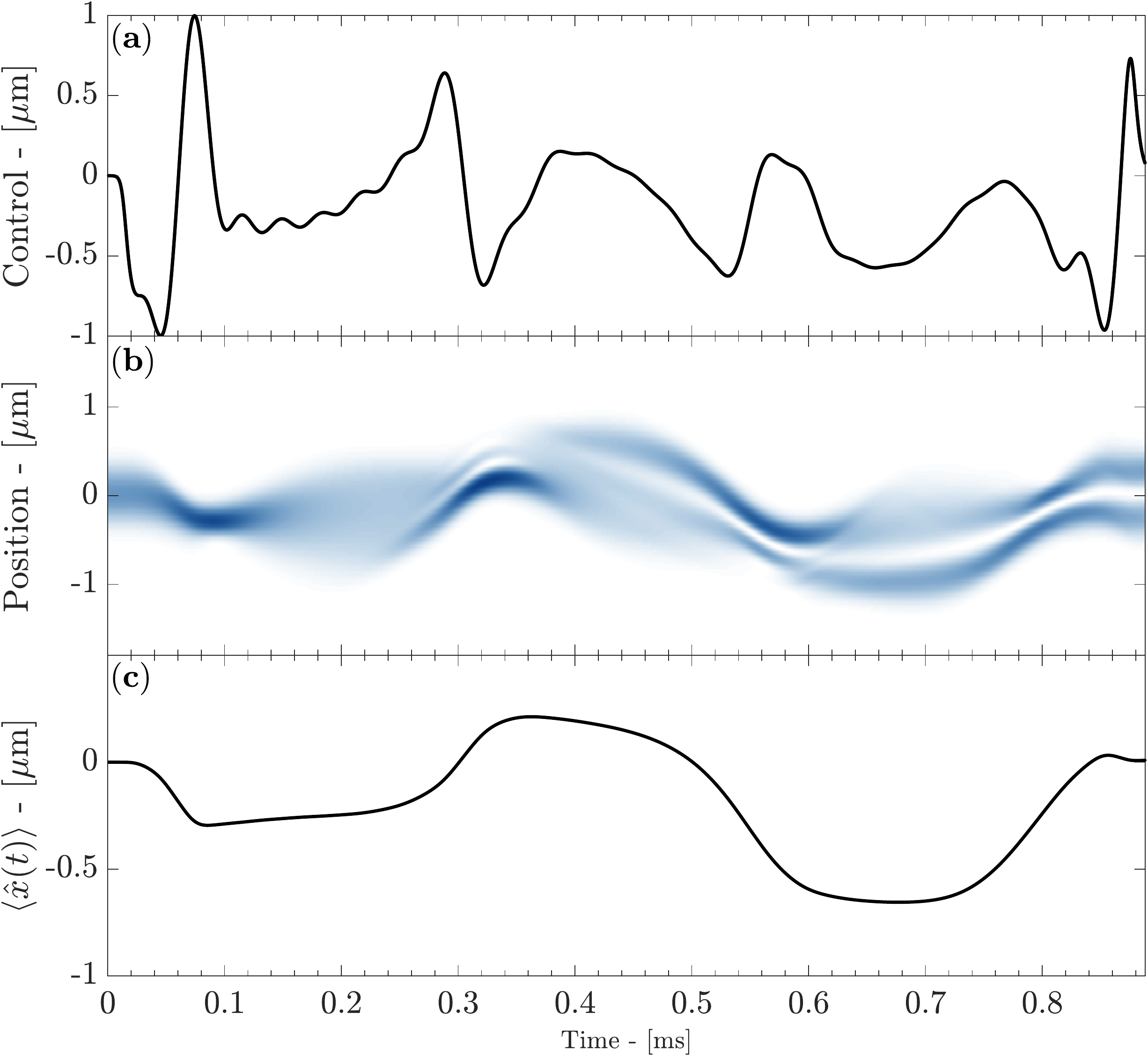}
\centering
\caption{(\textbf{a}) The fastest optimal control for Condensate Driving (CD) found using Global-Local optimization. (\textbf{b}) The density for the condensate when propagated along the optimal control for comparison with the control. (\textbf{c}) The expectation value of the position as a function of time ($\langle \hat{x}(t) \rangle$). This expectation value has a clear main oscillation component, which is close to the difference between the ground state and the first excited state.}
\label{fig:antonioOptimalControl}
\end{minipage}
\quad
\begin{minipage}[t]{.48\textwidth}
\includegraphics[width=\textwidth]{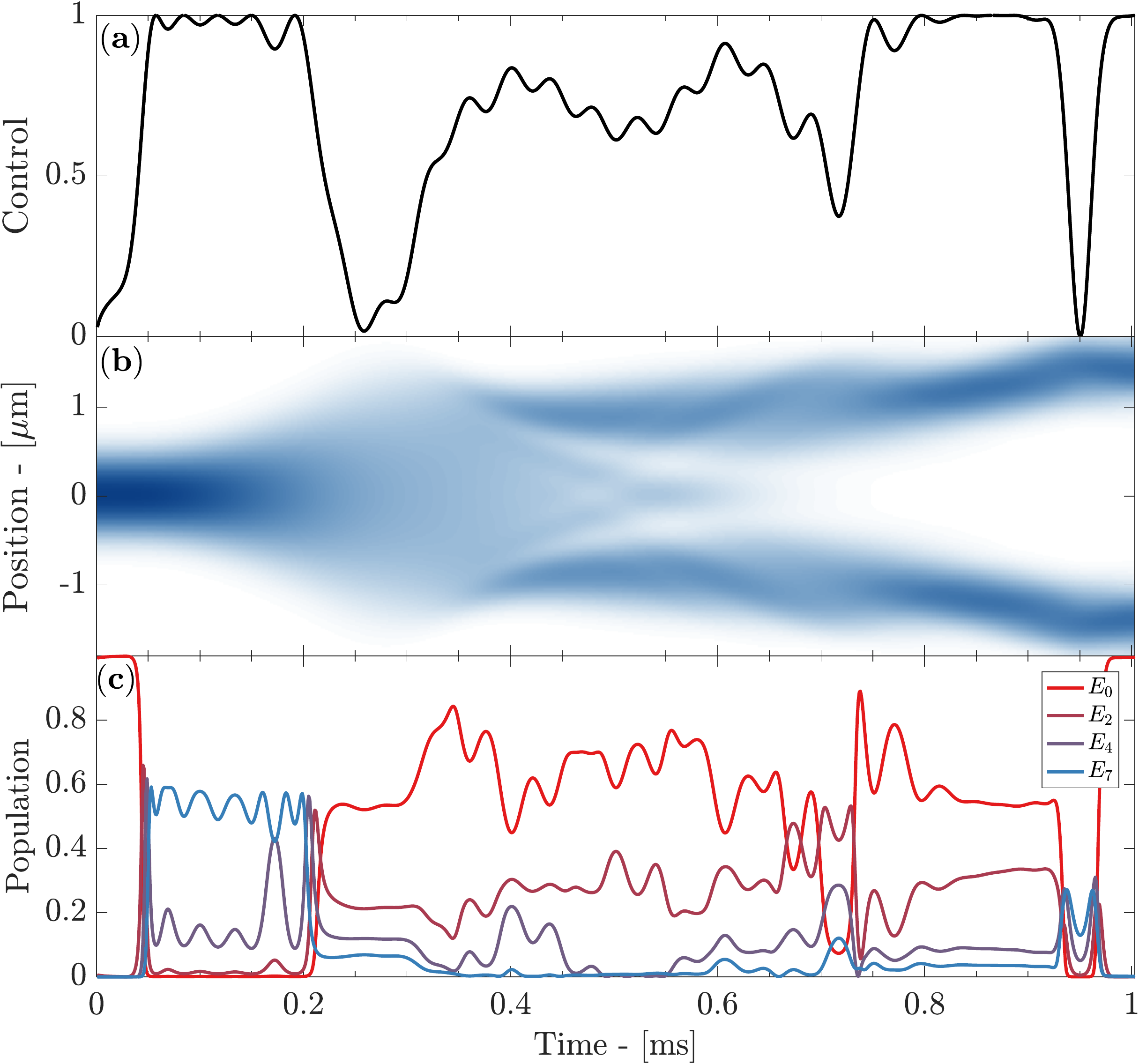}
\centering
\caption{(\textbf{a}) The fastest optimal control for Condensate Splitting (CS) found using Global-Local optimization. (\textbf{b}) The condensate's density when propagated along the control. (\textbf{c}) The population in the instantaneous energy eigenstates as a function of time. Only the even levels are populated due to the symmetry of the initial state and the potential.}
\label{fig:LesanovskyOptimalControl}
\end{minipage}
\end{figure*}

The result of our Global-Local algorithm on the CD-problem is presented in Fig. \ref{fig:antonioCondensate}\textbf{b}. The resulting numerical estimate for CD is $T_{\text{QSL}}^{\text{num}}=0.89\text{ms}$, which is the shortest duration with $F\geq 0.99$. This result is lower than the $1.09\text{ms}$ report in Ref. \cite{van2016optimal} where traditional multistarting and gradient-free optimization was used. The Global-Local algorithm also finds better fidelities than Ref. \cite{van2016optimal} below the estimated QSL ($T\leq T_{\text{QSL}}^{\text{num}}$) - see the blue dots in Fig. \ref{fig:antonioCondensate}\textbf{b}. This highlights that the landscape has become highly complex due to the strong duration constraint and traditional multistarting is no longer sufficient. In Ref. \cite{van2016optimal} a double $\sin^2$-behavior is found, which was interpreted to indicate that the solutions had mapped out the true QSL. Surprisingly, the Global-Local algorithm breaks this QSL and finds a new $F(T)$-curve that does not follow a $\sin^2$-behavior. This indicates that a $\sin^2$-behavior does not always imply, that the numerical results have identified the QSL. The Global-Local algorithm will theoretically identify the true QSL in the infinite time limit, due to our finite optimization time even better solutions could possibly exist. Therefore we cannot exclude the existence of a better $F(T)$-curve, which could have a $\sin^2$-behavior. Previous $\sin^2$-results from e.g. \cite{caneva2009optimal, caneva2011speeding} were all for a linear Schr\"{o}dinger equation, so it is not directly clear that these results generalize to the nonlinear dynamics studied here. The $F(T)$-curve has a distinct kink around $T=0.2 \text{ms}$. At short durations below $T_{\text{QSL}}^{\text{num}}$ the best control only displaces the condensate and at longer durations the control does a partial transfer of the wavefunction into the first excited state - see the insert in  Fig. \ref{fig:antonioCondensate}\textbf{b}. These two processes scale differently with respect to \textit{T}, so at some durations a displacement is better than a partial transfer. This gives an explanation for the kink in Fig. \ref{fig:antonioCondensate}\textbf{b}. A similar kink was observed in Ref. \cite{van2016optimal}. The optimal control curve and $\langle \hat{x}(t)\rangle$ is shown in Fig. \ref{fig:antonioOptimalControl}. The control is highly complex. However, a Fourier transformation of $\langle \hat{x}(t)\rangle$ reveals that the main oscillation frequency in $\langle \hat{x}(t)\rangle$ is close to the energy difference between the ground state and the first excited state, so the resulting control can partially be understood as resonant driving.

\subsection{Condensate Splitting}
In Condensate Splitting (CS) a BEC is split into two seperate BECs with the same phase. This splitting procedure can be used in a Mach-Zehnder type interferometer for matter waves \cite{schumm2005matter}. The quality of this interferometer depends on the quality of the splitting, which is optimized. The atoms are trapped in an Ioffe-Pritchard field configuration on the atom chip. The potential is created by applying RF-dressing, which causes mixing of the Zeeman levels of the $F=1$ manifold coupling them to the dressed states \cite{bucker2013vibrational}. In Ref. \cite{lesanovsky2006adiabatic} it is shown within the rotating wave approximation that this gives the potential
\begin{equation*}
    V(x,u(t))=g_F\mu_B \sqrt{\biggl(B_S(x)-\frac{\hbar \omega}{g_F\mu_B}\biggr)^2+\biggl(\frac{B_{\text{RF}}B_I}{2B_S(x)}\biggr)^2},
\end{equation*}
where $g_F$ is the \textit{g}-factor, $\mu_B$ is the Bohr magneton. $\omega = 1.26 \cdot 2\pi \text{MHz}$ is the field detuning. $B_I=1.0 \text{G}$ and $B_\text{RF}=(0.5+0.3u(t))\text{G}$ are a magnetic field component related to the inhomogenous Ioffe field and the experimentally adjustable RF-field. $B_S^2(x)=(Gx)^2+B_I^2$ is a static field where \textit{G} is the gradient of the Ioffe-field trap\cite{lesanovsky2006adiabatic,hohenester2007optimal,bucker2013vibrational}. For $u=0$ the potential is a single well and it dynamically changes into a double well as \textit{u} changes into $u=1$, which is shown in Fig. \ref{fig:lesanovskyCondensate}\textbf{a}.

The result of our Global-Local optimization on the CS-problem is presented in Fig. \ref{fig:lesanovskyCondensate}\textbf{b}. The resulting numerical estimate for
CS is $T_{\text{QSL}}^{\text{num}}=1.0\text{ms}$. To our knowledge there is no $F(T)$-curve in the literature for comparison but our results are faster than the $2.0\text{ms}$ report in Ref. \cite{jager2014optimal,hohenester2007optimal}. Neither in the case of CS do we find a $\sin^2$-behavior as seen on Fig. \ref{fig:lesanovskyCondensate}\textbf{b}. The optimal control curve and the population in the instantaneous linear energy eigenstates are shown in Fig. \ref{fig:LesanovskyOptimalControl}. The optimal control first excites the condensate as much as possible by applying the maximally allowed splitting for the first 1/5 of the control duration. This excites the BEC and by Heisenberg's time-energy uncertainty relation allows for fast motion in the Hilbert Space. The importance of constraints on the QSL is especially clear here, since the excitation process could be completed faster if a larger double-well splitting was allowed. After the initial excitation period the condensate is transferred into the target state. This optimal control is highly diabatic and clearly differs from adiabatically inspired solutions where the splitting is gradually turned on. The control found here is quite different from that in e.g. Ref \cite{jager2014optimal}. This highlights the Global-Local algorithm's ability to search control subspaces far from the adiabatic regime, which is important when approaching the QSL where the optimization landscape is complex.

\section{Learning} \label{sec:Learning}
\begin{figure}[t]
\includegraphics[width=1.00\columnwidth]{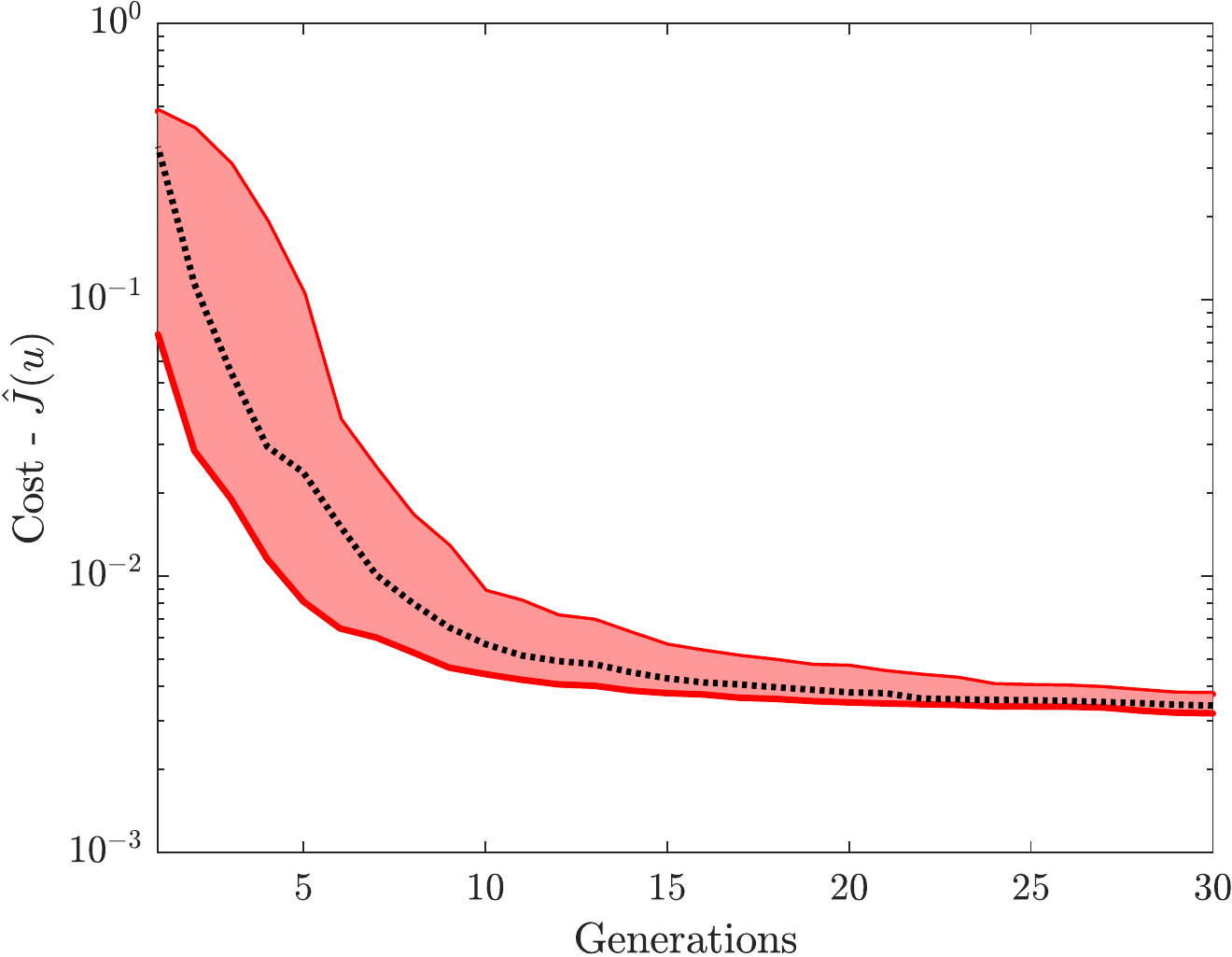}
\centering
\caption{Learning in the Global-Local algorithm in CS at $T_{\text{QSL}}^{\text{num}}=1.0\text{ms}$. The cost is shown as a function of the generations for the population. The dotted line is the median infidelity in the population and the shaded area shows the 25\%- and 75\%-quartiles.
}
\label{fig:learning}
\end{figure}
Finally we discuss the learning in the Global-Local optimization. In Fig. \ref{fig:learning} the distribution of cost values within each generation is shown as a function of generations in the optimization in CS at $T_{\text{QSL}}^{\text{num}}=1.0\text{ms}$. The figure shows that the Global-Local algorithm gradually decreases the median infidelity and thus learns a better solution strategy. In comparison, a multistarting algorithm would have a constant distribution set by the seeding strategy. We also did a traditional multistarting on the seeds in the initial population and none of the solutions achieved a similar value of the cost. The initial population is wide and it narrows as a function of generations. This shows that the optimization does an early exploration phase and subsequently spends successive generations on refining the current best members. However, this plot also suggest that the learning in the later stages of the algorithm could be substantially improved. With these early results we also cannot rule out that with sufficient fine tuning of the local optimization algorithms it might be possible to achieve similar results. The relative merit of improving local and global search methodologies represents an interesting avenue of future research.

\section{Conclusion}
We have implemented a combined a Global-Local optimization algorithm that combines evolutionary algorithms like \textsc{de} with local quantum control methods like \textsc{group}. This combination has allowed the improvement of existing estimates of the QSL in problems related to control of BECs. 

This scheme is directly applicable to other problems in quantum control where it might also give improvements in the QSL.

It would be possible to further tune the balance between exploration and exploitation by modifying the \textsc{DE} algorithm or using another explorative algorithm like CMA-ES. Within \textsc{DE}, exploration can be promoted by changing the scheme for the generation of the donor vector using a method like SaDE \cite{das2011differential}. Finally, it would also be very interesting to do a more in depth comparison with multistarting in order to better quantify the advantages of the Global-Local algorithm.

\section{Acknowledgements}
This work has funded by the European Research Council and the Lundbeck Foundation. We would also like to thank Jesper H. M. Jensen for useful discussions.

\bibliographystyle{apsrev4-1}
\bibliography{references.bib}

%merlin.mbs apsrev4-1.bst 2010-07-25 4.21a (PWD, AO, DPC) hacked
%Control: key (0)
%Control: author (72) initials jnrlst
%Control: editor formatted (1) identically to author
%Control: production of article title (-1) disabled
%Control: page (0) single
%Control: year (1) truncated
%Control: production of eprint (0) enabled
\begin{thebibliography}{39}%
\makeatletter
\providecommand \@ifxundefined [1]{%
 \@ifx{#1\undefined}
}%
\providecommand \@ifnum [1]{%
 \ifnum #1\expandafter \@firstoftwo
 \else \expandafter \@secondoftwo
 \fi
}%
\providecommand \@ifx [1]{%
 \ifx #1\expandafter \@firstoftwo
 \else \expandafter \@secondoftwo
 \fi
}%
\providecommand \natexlab [1]{#1}%
\providecommand \enquote  [1]{``#1''}%
\providecommand \bibnamefont  [1]{#1}%
\providecommand \bibfnamefont [1]{#1}%
\providecommand \citenamefont [1]{#1}%
\providecommand \href@noop [0]{\@secondoftwo}%
\providecommand \href [0]{\begingroup \@sanitize@url \@href}%
\providecommand \@href[1]{\@@startlink{#1}\@@href}%
\providecommand \@@href[1]{\endgroup#1\@@endlink}%
\providecommand \@sanitize@url [0]{\catcode `\\12\catcode `\$12\catcode
  `\&12\catcode `\#12\catcode `\^12\catcode `\_12\catcode `\%12\relax}%
\providecommand \@@startlink[1]{}%
\providecommand \@@endlink[0]{}%
\providecommand \url  [0]{\begingroup\@sanitize@url \@url }%
\providecommand \@url [1]{\endgroup\@href {#1}{\urlprefix }}%
\providecommand \urlprefix  [0]{URL }%
\providecommand \Eprint [0]{\href }%
\providecommand \doibase [0]{http://dx.doi.org/}%
\providecommand \selectlanguage [0]{\@gobble}%
\providecommand \bibinfo  [0]{\@secondoftwo}%
\providecommand \bibfield  [0]{\@secondoftwo}%
\providecommand \translation [1]{[#1]}%
\providecommand \BibitemOpen [0]{}%
\providecommand \bibitemStop [0]{}%
\providecommand \bibitemNoStop [0]{.\EOS\space}%
\providecommand \EOS [0]{\spacefactor3000\relax}%
\providecommand \BibitemShut  [1]{\csname bibitem#1\endcsname}%
\let\auto@bib@innerbib\@empty
%</preamble>
\bibitem [{\citenamefont {Assion}\ \emph {et~al.}(1998)\citenamefont {Assion},
  \citenamefont {Baumert}, \citenamefont {Bergt}, \citenamefont {Brixner},
  \citenamefont {Kiefer}, \citenamefont {Seyfried}, \citenamefont {Strehle},\
  and\ \citenamefont {Gerber}}]{assion1998control}%
  \BibitemOpen
  \bibfield  {author} {\bibinfo {author} {\bibfnamefont {A.}~\bibnamefont
  {Assion}}, \bibinfo {author} {\bibfnamefont {T.}~\bibnamefont {Baumert}},
  \bibinfo {author} {\bibfnamefont {M.}~\bibnamefont {Bergt}}, \bibinfo
  {author} {\bibfnamefont {T.}~\bibnamefont {Brixner}}, \bibinfo {author}
  {\bibfnamefont {B.}~\bibnamefont {Kiefer}}, \bibinfo {author} {\bibfnamefont
  {V.}~\bibnamefont {Seyfried}}, \bibinfo {author} {\bibfnamefont
  {M.}~\bibnamefont {Strehle}}, \ and\ \bibinfo {author} {\bibfnamefont
  {G.}~\bibnamefont {Gerber}},\ }\href@noop {} {\bibfield  {journal} {\bibinfo
  {journal} {Science}\ }\textbf {\bibinfo {volume} {282}},\ \bibinfo {pages}
  {919} (\bibinfo {year} {1998})}\BibitemShut {NoStop}%
\bibitem [{\citenamefont {Meshulach}\ and\ \citenamefont
  {Silberberg}(1998)}]{meshulach1998coherent}%
  \BibitemOpen
  \bibfield  {author} {\bibinfo {author} {\bibfnamefont {D.}~\bibnamefont
  {Meshulach}}\ and\ \bibinfo {author} {\bibfnamefont {Y.}~\bibnamefont
  {Silberberg}},\ }\href@noop {} {\bibfield  {journal} {\bibinfo  {journal}
  {Nature}\ }\textbf {\bibinfo {volume} {396}},\ \bibinfo {pages} {239}
  (\bibinfo {year} {1998})}\BibitemShut {NoStop}%
\bibitem [{\citenamefont {Schulte-Herbr{\"u}ggen}\ \emph
  {et~al.}(2005)\citenamefont {Schulte-Herbr{\"u}ggen}, \citenamefont
  {Sp{\"o}rl}, \citenamefont {Khaneja},\ and\ \citenamefont
  {Glaser}}]{schulte2005optimal}%
  \BibitemOpen
  \bibfield  {author} {\bibinfo {author} {\bibfnamefont {T.}~\bibnamefont
  {Schulte-Herbr{\"u}ggen}}, \bibinfo {author} {\bibfnamefont {A.}~\bibnamefont
  {Sp{\"o}rl}}, \bibinfo {author} {\bibfnamefont {N.}~\bibnamefont {Khaneja}},
  \ and\ \bibinfo {author} {\bibfnamefont {S.}~\bibnamefont {Glaser}},\
  }\href@noop {} {\bibfield  {journal} {\bibinfo  {journal} {Physical Review
  A}\ }\textbf {\bibinfo {volume} {72}},\ \bibinfo {pages} {042331} (\bibinfo
  {year} {2005})}\BibitemShut {NoStop}%
\bibitem [{\citenamefont {Doria}\ \emph {et~al.}(2011)\citenamefont {Doria},
  \citenamefont {Calarco},\ and\ \citenamefont
  {Montangero}}]{doria2011optimal}%
  \BibitemOpen
  \bibfield  {author} {\bibinfo {author} {\bibfnamefont {P.}~\bibnamefont
  {Doria}}, \bibinfo {author} {\bibfnamefont {T.}~\bibnamefont {Calarco}}, \
  and\ \bibinfo {author} {\bibfnamefont {S.}~\bibnamefont {Montangero}},\
  }\href@noop {} {\bibfield  {journal} {\bibinfo  {journal} {Physical review
  letters}\ }\textbf {\bibinfo {volume} {106}},\ \bibinfo {pages} {190501}
  (\bibinfo {year} {2011})}\BibitemShut {NoStop}%
\bibitem [{\citenamefont {van Frank}\ \emph {et~al.}(2016)\citenamefont {van
  Frank}, \citenamefont {Bonneau}, \citenamefont {Schmiedmayer}, \citenamefont
  {Hild}, \citenamefont {Gross}, \citenamefont {Cheneau}, \citenamefont
  {Bloch}, \citenamefont {Pichler}, \citenamefont {Negretti}, \citenamefont
  {Calarco} \emph {et~al.}}]{van2016optimal}%
  \BibitemOpen
  \bibfield  {author} {\bibinfo {author} {\bibfnamefont {S.}~\bibnamefont {van
  Frank}}, \bibinfo {author} {\bibfnamefont {M.}~\bibnamefont {Bonneau}},
  \bibinfo {author} {\bibfnamefont {J.}~\bibnamefont {Schmiedmayer}}, \bibinfo
  {author} {\bibfnamefont {S.}~\bibnamefont {Hild}}, \bibinfo {author}
  {\bibfnamefont {C.}~\bibnamefont {Gross}}, \bibinfo {author} {\bibfnamefont
  {M.}~\bibnamefont {Cheneau}}, \bibinfo {author} {\bibfnamefont
  {I.}~\bibnamefont {Bloch}}, \bibinfo {author} {\bibfnamefont
  {T.}~\bibnamefont {Pichler}}, \bibinfo {author} {\bibfnamefont
  {A.}~\bibnamefont {Negretti}}, \bibinfo {author} {\bibfnamefont
  {T.}~\bibnamefont {Calarco}},  \emph {et~al.},\ }\href@noop {} {\bibfield
  {journal} {\bibinfo  {journal} {Scientific reports}\ }\textbf {\bibinfo
  {volume} {6}} (\bibinfo {year} {2016})}\BibitemShut {NoStop}%
\bibitem [{\citenamefont {Levitin}\ and\ \citenamefont
  {Toffoli}(2009)}]{levitin2009fundamental}%
  \BibitemOpen
  \bibfield  {author} {\bibinfo {author} {\bibfnamefont {L.~B.}\ \bibnamefont
  {Levitin}}\ and\ \bibinfo {author} {\bibfnamefont {T.}~\bibnamefont
  {Toffoli}},\ }\href@noop {} {\bibfield  {journal} {\bibinfo  {journal}
  {Physical review letters}\ }\textbf {\bibinfo {volume} {103}},\ \bibinfo
  {pages} {160502} (\bibinfo {year} {2009})}\BibitemShut {NoStop}%
\bibitem [{\citenamefont {Taddei}\ \emph {et~al.}(2013)\citenamefont {Taddei},
  \citenamefont {Escher}, \citenamefont {Davidovich},\ and\ \citenamefont
  {de~Matos~Filho}}]{taddei2013quantum}%
  \BibitemOpen
  \bibfield  {author} {\bibinfo {author} {\bibfnamefont {M.~M.}\ \bibnamefont
  {Taddei}}, \bibinfo {author} {\bibfnamefont {B.~M.}\ \bibnamefont {Escher}},
  \bibinfo {author} {\bibfnamefont {L.}~\bibnamefont {Davidovich}}, \ and\
  \bibinfo {author} {\bibfnamefont {R.~L.}\ \bibnamefont {de~Matos~Filho}},\
  }\href@noop {} {\bibfield  {journal} {\bibinfo  {journal} {Physical review
  letters}\ }\textbf {\bibinfo {volume} {110}},\ \bibinfo {pages} {050402}
  (\bibinfo {year} {2013})}\BibitemShut {NoStop}%
\bibitem [{\citenamefont {Caneva}\ \emph {et~al.}(2009)\citenamefont {Caneva},
  \citenamefont {Murphy}, \citenamefont {Calarco}, \citenamefont {Fazio},
  \citenamefont {Montangero}, \citenamefont {Giovannetti},\ and\ \citenamefont
  {Santoro}}]{caneva2009optimal}%
  \BibitemOpen
  \bibfield  {author} {\bibinfo {author} {\bibfnamefont {T.}~\bibnamefont
  {Caneva}}, \bibinfo {author} {\bibfnamefont {M.}~\bibnamefont {Murphy}},
  \bibinfo {author} {\bibfnamefont {T.}~\bibnamefont {Calarco}}, \bibinfo
  {author} {\bibfnamefont {R.}~\bibnamefont {Fazio}}, \bibinfo {author}
  {\bibfnamefont {S.}~\bibnamefont {Montangero}}, \bibinfo {author}
  {\bibfnamefont {V.}~\bibnamefont {Giovannetti}}, \ and\ \bibinfo {author}
  {\bibfnamefont {G.~E.}\ \bibnamefont {Santoro}},\ }\href@noop {} {\bibfield
  {journal} {\bibinfo  {journal} {Physical review letters}\ }\textbf {\bibinfo
  {volume} {103}},\ \bibinfo {pages} {240501} (\bibinfo {year}
  {2009})}\BibitemShut {NoStop}%
\bibitem [{\citenamefont {Deffner}\ and\ \citenamefont
  {Campbell}(2017)}]{deffner2017quantum}%
  \BibitemOpen
  \bibfield  {author} {\bibinfo {author} {\bibfnamefont {S.}~\bibnamefont
  {Deffner}}\ and\ \bibinfo {author} {\bibfnamefont {S.}~\bibnamefont
  {Campbell}},\ }\href {http://stacks.iop.org/1751-8121/50/i=45/a=453001}
  {\bibfield  {journal} {\bibinfo  {journal} {Journal of Physics A:
  Mathematical and Theoretical}\ }\textbf {\bibinfo {volume} {50}},\ \bibinfo
  {pages} {453001} (\bibinfo {year} {2017})}\BibitemShut {NoStop}%
\bibitem [{\citenamefont {Werschnik}\ and\ \citenamefont
  {Gross}(2007)}]{werschnik2007quantum}%
  \BibitemOpen
  \bibfield  {author} {\bibinfo {author} {\bibfnamefont {J.}~\bibnamefont
  {Werschnik}}\ and\ \bibinfo {author} {\bibfnamefont {E.}~\bibnamefont
  {Gross}},\ }\href@noop {} {\bibfield  {journal} {\bibinfo  {journal} {Journal
  of Physics B: Atomic, Molecular and Optical Physics}\ }\textbf {\bibinfo
  {volume} {40}},\ \bibinfo {pages} {R175} (\bibinfo {year}
  {2007})}\BibitemShut {NoStop}%
\bibitem [{\citenamefont {Russell}\ \emph {et~al.}(2017)\citenamefont
  {Russell}, \citenamefont {Rabitz},\ and\ \citenamefont
  {Wu}}]{russell2016quantum}%
  \BibitemOpen
  \bibfield  {author} {\bibinfo {author} {\bibfnamefont {B.}~\bibnamefont
  {Russell}}, \bibinfo {author} {\bibfnamefont {H.}~\bibnamefont {Rabitz}}, \
  and\ \bibinfo {author} {\bibfnamefont {R.-B.}\ \bibnamefont {Wu}},\ }\href
  {http://stacks.iop.org/1751-8121/50/i=20/a=205302} {\bibfield  {journal}
  {\bibinfo  {journal} {Journal of Physics A: Mathematical and Theoretical}\
  }\textbf {\bibinfo {volume} {50}},\ \bibinfo {pages} {205302} (\bibinfo
  {year} {2017})}\BibitemShut {NoStop}%
\bibitem [{\citenamefont {Rabitz}\ \emph {et~al.}(2004)\citenamefont {Rabitz},
  \citenamefont {Hsieh},\ and\ \citenamefont {Rosenthal}}]{rabitz2004quantum}%
  \BibitemOpen
  \bibfield  {author} {\bibinfo {author} {\bibfnamefont {H.~A.}\ \bibnamefont
  {Rabitz}}, \bibinfo {author} {\bibfnamefont {M.~M.}\ \bibnamefont {Hsieh}}, \
  and\ \bibinfo {author} {\bibfnamefont {C.~M.}\ \bibnamefont {Rosenthal}},\
  }\href@noop {} {\bibfield  {journal} {\bibinfo  {journal} {Science}\ }\textbf
  {\bibinfo {volume} {303}},\ \bibinfo {pages} {1998} (\bibinfo {year}
  {2004})}\BibitemShut {NoStop}%
\bibitem [{\citenamefont {Zhdanov}\ and\ \citenamefont
  {Seideman}(2015)}]{zhdanov2015role}%
  \BibitemOpen
  \bibfield  {author} {\bibinfo {author} {\bibfnamefont {D.~V.}\ \bibnamefont
  {Zhdanov}}\ and\ \bibinfo {author} {\bibfnamefont {T.}~\bibnamefont
  {Seideman}},\ }\href@noop {} {\bibfield  {journal} {\bibinfo  {journal}
  {Physical Review A}\ }\textbf {\bibinfo {volume} {92}},\ \bibinfo {pages}
  {052109} (\bibinfo {year} {2015})}\BibitemShut {NoStop}%
\bibitem [{\citenamefont {Bukov}\ \emph {et~al.}(2017)\citenamefont {Bukov},
  \citenamefont {Day}, \citenamefont {Sels}, \citenamefont {Weinberg},
  \citenamefont {Polkovnikov},\ and\ \citenamefont {Mehta}}]{bukov2017machine}%
  \BibitemOpen
  \bibfield  {author} {\bibinfo {author} {\bibfnamefont {M.}~\bibnamefont
  {Bukov}}, \bibinfo {author} {\bibfnamefont {A.~G.}\ \bibnamefont {Day}},
  \bibinfo {author} {\bibfnamefont {D.}~\bibnamefont {Sels}}, \bibinfo {author}
  {\bibfnamefont {P.}~\bibnamefont {Weinberg}}, \bibinfo {author}
  {\bibfnamefont {A.}~\bibnamefont {Polkovnikov}}, \ and\ \bibinfo {author}
  {\bibfnamefont {P.}~\bibnamefont {Mehta}},\ }\href@noop {} {\bibfield
  {journal} {\bibinfo  {journal} {arXiv preprint arXiv:1705.00565}\ } (\bibinfo
  {year} {2017})}\BibitemShut {NoStop}%
\bibitem [{\citenamefont {Neri}\ and\ \citenamefont
  {Cotta}(2012)}]{neri2012memetic}%
  \BibitemOpen
  \bibfield  {author} {\bibinfo {author} {\bibfnamefont {F.}~\bibnamefont
  {Neri}}\ and\ \bibinfo {author} {\bibfnamefont {C.}~\bibnamefont {Cotta}},\
  }\href@noop {} {\bibfield  {journal} {\bibinfo  {journal} {Swarm and
  Evolutionary Computation}\ }\textbf {\bibinfo {volume} {2}},\ \bibinfo
  {pages} {1} (\bibinfo {year} {2012})}\BibitemShut {NoStop}%
\bibitem [{\citenamefont {S{\o}rensen}\ \emph {et~al.}(2018)\citenamefont
  {S{\o}rensen}, \citenamefont {Aranburu}, \citenamefont {Heinzel},\ and\
  \citenamefont {Sherson}}]{localPaper}%
  \BibitemOpen
  \bibfield  {author} {\bibinfo {author} {\bibfnamefont {J.~J. W.~H.}\
  \bibnamefont {S{\o}rensen}}, \bibinfo {author} {\bibfnamefont
  {M.}~\bibnamefont {Aranburu}}, \bibinfo {author} {\bibfnamefont
  {T.}~\bibnamefont {Heinzel}}, \ and\ \bibinfo {author} {\bibfnamefont
  {J.}~\bibnamefont {Sherson}},\ }\href@noop {} {\bibfield  {journal} {\bibinfo
   {journal} {To be published}\ } (\bibinfo {year} {2018})}\BibitemShut
  {NoStop}%
\bibitem [{\citenamefont {Khaneja}\ \emph {et~al.}(2005)\citenamefont
  {Khaneja}, \citenamefont {Reiss}, \citenamefont {Kehlet}, \citenamefont
  {Schulte-Herbr{\"u}ggen},\ and\ \citenamefont {Glaser}}]{khaneja2005optimal}%
  \BibitemOpen
  \bibfield  {author} {\bibinfo {author} {\bibfnamefont {N.}~\bibnamefont
  {Khaneja}}, \bibinfo {author} {\bibfnamefont {T.}~\bibnamefont {Reiss}},
  \bibinfo {author} {\bibfnamefont {C.}~\bibnamefont {Kehlet}}, \bibinfo
  {author} {\bibfnamefont {T.}~\bibnamefont {Schulte-Herbr{\"u}ggen}}, \ and\
  \bibinfo {author} {\bibfnamefont {S.~J.}\ \bibnamefont {Glaser}},\
  }\href@noop {} {\bibfield  {journal} {\bibinfo  {journal} {Journal of
  magnetic resonance}\ }\textbf {\bibinfo {volume} {172}},\ \bibinfo {pages}
  {296} (\bibinfo {year} {2005})}\BibitemShut {NoStop}%
\bibitem [{\citenamefont {Caneva}\ \emph
  {et~al.}(2011{\natexlab{a}})\citenamefont {Caneva}, \citenamefont {Calarco},\
  and\ \citenamefont {Montangero}}]{caneva2011chopped}%
  \BibitemOpen
  \bibfield  {author} {\bibinfo {author} {\bibfnamefont {T.}~\bibnamefont
  {Caneva}}, \bibinfo {author} {\bibfnamefont {T.}~\bibnamefont {Calarco}}, \
  and\ \bibinfo {author} {\bibfnamefont {S.}~\bibnamefont {Montangero}},\
  }\href@noop {} {\bibfield  {journal} {\bibinfo  {journal} {Physical Review
  A}\ }\textbf {\bibinfo {volume} {84}},\ \bibinfo {pages} {022326} (\bibinfo
  {year} {2011}{\natexlab{a}})}\BibitemShut {NoStop}%
\bibitem [{\citenamefont {Brouzos}\ \emph {et~al.}(2015)\citenamefont
  {Brouzos}, \citenamefont {Streltsov}, \citenamefont {Negretti}, \citenamefont
  {Said}, \citenamefont {Caneva}, \citenamefont {Montangero},\ and\
  \citenamefont {Calarco}}]{brouzos2015quantum}%
  \BibitemOpen
  \bibfield  {author} {\bibinfo {author} {\bibfnamefont {I.}~\bibnamefont
  {Brouzos}}, \bibinfo {author} {\bibfnamefont {A.~I.}\ \bibnamefont
  {Streltsov}}, \bibinfo {author} {\bibfnamefont {A.}~\bibnamefont {Negretti}},
  \bibinfo {author} {\bibfnamefont {R.~S.}\ \bibnamefont {Said}}, \bibinfo
  {author} {\bibfnamefont {T.}~\bibnamefont {Caneva}}, \bibinfo {author}
  {\bibfnamefont {S.}~\bibnamefont {Montangero}}, \ and\ \bibinfo {author}
  {\bibfnamefont {T.}~\bibnamefont {Calarco}},\ }\href@noop {} {\bibfield
  {journal} {\bibinfo  {journal} {Physical Review A}\ }\textbf {\bibinfo
  {volume} {92}},\ \bibinfo {pages} {062110} (\bibinfo {year}
  {2015})}\BibitemShut {NoStop}%
\bibitem [{\citenamefont {Gajdacz}\ \emph {et~al.}(2015)\citenamefont
  {Gajdacz}, \citenamefont {Das}, \citenamefont {Arlt}, \citenamefont
  {Sherson},\ and\ \citenamefont {Opatrn{\`y}}}]{gajdacz2015time}%
  \BibitemOpen
  \bibfield  {author} {\bibinfo {author} {\bibfnamefont {M.}~\bibnamefont
  {Gajdacz}}, \bibinfo {author} {\bibfnamefont {K.~K.}\ \bibnamefont {Das}},
  \bibinfo {author} {\bibfnamefont {J.}~\bibnamefont {Arlt}}, \bibinfo {author}
  {\bibfnamefont {J.~F.}\ \bibnamefont {Sherson}}, \ and\ \bibinfo {author}
  {\bibfnamefont {T.}~\bibnamefont {Opatrn{\`y}}},\ }\href@noop {} {\bibfield
  {journal} {\bibinfo  {journal} {Physical Review A}\ }\textbf {\bibinfo
  {volume} {92}},\ \bibinfo {pages} {062106} (\bibinfo {year}
  {2015})}\BibitemShut {NoStop}%
\bibitem [{\citenamefont {S{\o}rensen}\ \emph {et~al.}(2016)\citenamefont
  {S{\o}rensen}, \citenamefont {Pedersen}, \citenamefont {Munch}, \citenamefont
  {Haikka}, \citenamefont {Jensen}, \citenamefont {Planke}, \citenamefont
  {Andreasen}, \citenamefont {Gajdacz}, \citenamefont {M{\o}lmer},
  \citenamefont {Lieberoth} \emph {et~al.}}]{sorensen2016exploring}%
  \BibitemOpen
  \bibfield  {author} {\bibinfo {author} {\bibfnamefont {J.~J.~W.}\
  \bibnamefont {S{\o}rensen}}, \bibinfo {author} {\bibfnamefont {M.~K.}\
  \bibnamefont {Pedersen}}, \bibinfo {author} {\bibfnamefont {M.}~\bibnamefont
  {Munch}}, \bibinfo {author} {\bibfnamefont {P.}~\bibnamefont {Haikka}},
  \bibinfo {author} {\bibfnamefont {J.~H.}\ \bibnamefont {Jensen}}, \bibinfo
  {author} {\bibfnamefont {T.}~\bibnamefont {Planke}}, \bibinfo {author}
  {\bibfnamefont {M.~G.}\ \bibnamefont {Andreasen}}, \bibinfo {author}
  {\bibfnamefont {M.}~\bibnamefont {Gajdacz}}, \bibinfo {author} {\bibfnamefont
  {K.}~\bibnamefont {M{\o}lmer}}, \bibinfo {author} {\bibfnamefont
  {A.}~\bibnamefont {Lieberoth}},  \emph {et~al.},\ }\href@noop {} {\bibfield
  {journal} {\bibinfo  {journal} {Nature}\ }\textbf {\bibinfo {volume} {532}},\
  \bibinfo {pages} {210} (\bibinfo {year} {2016})}\BibitemShut {NoStop}%
\bibitem [{\citenamefont {Schirmer}\ and\ \citenamefont
  {de~Fouquieres}(2011)}]{schirmer2011efficient}%
  \BibitemOpen
  \bibfield  {author} {\bibinfo {author} {\bibfnamefont {S.~G.}\ \bibnamefont
  {Schirmer}}\ and\ \bibinfo {author} {\bibfnamefont {P.}~\bibnamefont
  {de~Fouquieres}},\ }\href@noop {} {\bibfield  {journal} {\bibinfo  {journal}
  {New Journal of Physics}\ }\textbf {\bibinfo {volume} {13}},\ \bibinfo
  {pages} {073029} (\bibinfo {year} {2011})}\BibitemShut {NoStop}%
\bibitem [{\citenamefont {Hohenester}\ \emph {et~al.}(2007)\citenamefont
  {Hohenester}, \citenamefont {Rekdal}, \citenamefont {Borz{\`\i}},\ and\
  \citenamefont {Schmiedmayer}}]{hohenester2007optimal}%
  \BibitemOpen
  \bibfield  {author} {\bibinfo {author} {\bibfnamefont {U.}~\bibnamefont
  {Hohenester}}, \bibinfo {author} {\bibfnamefont {P.~K.}\ \bibnamefont
  {Rekdal}}, \bibinfo {author} {\bibfnamefont {A.}~\bibnamefont {Borz{\`\i}}},
  \ and\ \bibinfo {author} {\bibfnamefont {J.}~\bibnamefont {Schmiedmayer}},\
  }\href@noop {} {\bibfield  {journal} {\bibinfo  {journal} {Physical Review
  A}\ }\textbf {\bibinfo {volume} {75}},\ \bibinfo {pages} {023602} (\bibinfo
  {year} {2007})}\BibitemShut {NoStop}%
\bibitem [{\citenamefont {Mennemann}\ \emph {et~al.}(2015)\citenamefont
  {Mennemann}, \citenamefont {Matthes}, \citenamefont {Weish{\"a}upl},\ and\
  \citenamefont {Langen}}]{mennemann2015optimal}%
  \BibitemOpen
  \bibfield  {author} {\bibinfo {author} {\bibfnamefont {J.-F.}\ \bibnamefont
  {Mennemann}}, \bibinfo {author} {\bibfnamefont {D.}~\bibnamefont {Matthes}},
  \bibinfo {author} {\bibfnamefont {R.-M.}\ \bibnamefont {Weish{\"a}upl}}, \
  and\ \bibinfo {author} {\bibfnamefont {T.}~\bibnamefont {Langen}},\
  }\href@noop {} {\bibfield  {journal} {\bibinfo  {journal} {New Journal of
  Physics}\ }\textbf {\bibinfo {volume} {17}},\ \bibinfo {pages} {113027}
  (\bibinfo {year} {2015})}\BibitemShut {NoStop}%
\bibitem [{\citenamefont {Lloyd}\ and\ \citenamefont
  {Montangero}(2014)}]{lloyd2014information}%
  \BibitemOpen
  \bibfield  {author} {\bibinfo {author} {\bibfnamefont {S.}~\bibnamefont
  {Lloyd}}\ and\ \bibinfo {author} {\bibfnamefont {S.}~\bibnamefont
  {Montangero}},\ }\href@noop {} {\bibfield  {journal} {\bibinfo  {journal}
  {Physical review letters}\ }\textbf {\bibinfo {volume} {113}},\ \bibinfo
  {pages} {010502} (\bibinfo {year} {2014})}\BibitemShut {NoStop}%
\bibitem [{\citenamefont {J{\"a}ger}\ \emph {et~al.}(2014)\citenamefont
  {J{\"a}ger}, \citenamefont {Reich}, \citenamefont {Goerz}, \citenamefont
  {Koch},\ and\ \citenamefont {Hohenester}}]{jager2014optimal}%
  \BibitemOpen
  \bibfield  {author} {\bibinfo {author} {\bibfnamefont {G.}~\bibnamefont
  {J{\"a}ger}}, \bibinfo {author} {\bibfnamefont {D.~M.}\ \bibnamefont
  {Reich}}, \bibinfo {author} {\bibfnamefont {M.~H.}\ \bibnamefont {Goerz}},
  \bibinfo {author} {\bibfnamefont {C.~P.}\ \bibnamefont {Koch}}, \ and\
  \bibinfo {author} {\bibfnamefont {U.}~\bibnamefont {Hohenester}},\
  }\href@noop {} {\bibfield  {journal} {\bibinfo  {journal} {Physical Review
  A}\ }\textbf {\bibinfo {volume} {90}},\ \bibinfo {pages} {033628} (\bibinfo
  {year} {2014})}\BibitemShut {NoStop}%
\bibitem [{\citenamefont {Ugray}\ \emph {et~al.}(2007)\citenamefont {Ugray},
  \citenamefont {Lasdon}, \citenamefont {Plummer}, \citenamefont {Glover},
  \citenamefont {Kelly},\ and\ \citenamefont {Mart{\'\i}}}]{ugray2007scatter}%
  \BibitemOpen
  \bibfield  {author} {\bibinfo {author} {\bibfnamefont {Z.}~\bibnamefont
  {Ugray}}, \bibinfo {author} {\bibfnamefont {L.}~\bibnamefont {Lasdon}},
  \bibinfo {author} {\bibfnamefont {J.}~\bibnamefont {Plummer}}, \bibinfo
  {author} {\bibfnamefont {F.}~\bibnamefont {Glover}}, \bibinfo {author}
  {\bibfnamefont {J.}~\bibnamefont {Kelly}}, \ and\ \bibinfo {author}
  {\bibfnamefont {R.}~\bibnamefont {Mart{\'\i}}},\ }\href@noop {} {\bibfield
  {journal} {\bibinfo  {journal} {INFORMS Journal on Computing}\ }\textbf
  {\bibinfo {volume} {19}},\ \bibinfo {pages} {328} (\bibinfo {year}
  {2007})}\BibitemShut {NoStop}%
\bibitem [{\citenamefont {Zahedinejad}\ \emph {et~al.}(2015)\citenamefont
  {Zahedinejad}, \citenamefont {Ghosh},\ and\ \citenamefont
  {Sanders}}]{zahedinejad2015high}%
  \BibitemOpen
  \bibfield  {author} {\bibinfo {author} {\bibfnamefont {E.}~\bibnamefont
  {Zahedinejad}}, \bibinfo {author} {\bibfnamefont {J.}~\bibnamefont {Ghosh}},
  \ and\ \bibinfo {author} {\bibfnamefont {B.~C.}\ \bibnamefont {Sanders}},\
  }\href@noop {} {\bibfield  {journal} {\bibinfo  {journal} {Physical review
  letters}\ }\textbf {\bibinfo {volume} {114}},\ \bibinfo {pages} {200502}
  (\bibinfo {year} {2015})}\BibitemShut {NoStop}%
\bibitem [{\citenamefont {Zahedinejad}\ \emph {et~al.}(2014)\citenamefont
  {Zahedinejad}, \citenamefont {Schirmer},\ and\ \citenamefont
  {Sanders}}]{zahedinejad2014evolutionary}%
  \BibitemOpen
  \bibfield  {author} {\bibinfo {author} {\bibfnamefont {E.}~\bibnamefont
  {Zahedinejad}}, \bibinfo {author} {\bibfnamefont {S.}~\bibnamefont
  {Schirmer}}, \ and\ \bibinfo {author} {\bibfnamefont {B.~C.}\ \bibnamefont
  {Sanders}},\ }\href@noop {} {\bibfield  {journal} {\bibinfo  {journal}
  {Physical Review A}\ }\textbf {\bibinfo {volume} {90}},\ \bibinfo {pages}
  {032310} (\bibinfo {year} {2014})}\BibitemShut {NoStop}%
\bibitem [{\citenamefont {Kennedy}(2011)}]{kennedy2011particle}%
  \BibitemOpen
  \bibfield  {author} {\bibinfo {author} {\bibfnamefont {J.}~\bibnamefont
  {Kennedy}},\ }in\ \href@noop {} {\emph {\bibinfo {booktitle} {Encyclopedia of
  machine learning}}}\ (\bibinfo  {publisher} {Springer},\ \bibinfo {year}
  {2011})\ pp.\ \bibinfo {pages} {760--766}\BibitemShut {NoStop}%
\bibitem [{\citenamefont {Kirkpatrick}\ \emph {et~al.}(1983)\citenamefont
  {Kirkpatrick}, \citenamefont {Gelatt}, \citenamefont {Vecchi} \emph
  {et~al.}}]{kirkpatrick1983optimization}%
  \BibitemOpen
  \bibfield  {author} {\bibinfo {author} {\bibfnamefont {S.}~\bibnamefont
  {Kirkpatrick}}, \bibinfo {author} {\bibfnamefont {C.~D.}\ \bibnamefont
  {Gelatt}}, \bibinfo {author} {\bibfnamefont {M.~P.}\ \bibnamefont {Vecchi}},
  \emph {et~al.},\ }\href@noop {} {\bibfield  {journal} {\bibinfo  {journal}
  {science}\ }\textbf {\bibinfo {volume} {220}},\ \bibinfo {pages} {671}
  (\bibinfo {year} {1983})}\BibitemShut {NoStop}%
\bibitem [{\citenamefont {Das}\ and\ \citenamefont
  {Suganthan}(2011)}]{das2011differential}%
  \BibitemOpen
  \bibfield  {author} {\bibinfo {author} {\bibfnamefont {S.}~\bibnamefont
  {Das}}\ and\ \bibinfo {author} {\bibfnamefont {P.~N.}\ \bibnamefont
  {Suganthan}},\ }\href@noop {} {\bibfield  {journal} {\bibinfo  {journal}
  {IEEE transactions on evolutionary computation}\ }\textbf {\bibinfo {volume}
  {15}},\ \bibinfo {pages} {4} (\bibinfo {year} {2011})}\BibitemShut {NoStop}%
\bibitem [{\citenamefont {Hansen}\ \emph {et~al.}(2003)\citenamefont {Hansen},
  \citenamefont {M{\"u}ller},\ and\ \citenamefont
  {Koumoutsakos}}]{hansen2003reducing}%
  \BibitemOpen
  \bibfield  {author} {\bibinfo {author} {\bibfnamefont {N.}~\bibnamefont
  {Hansen}}, \bibinfo {author} {\bibfnamefont {S.~D.}\ \bibnamefont
  {M{\"u}ller}}, \ and\ \bibinfo {author} {\bibfnamefont {P.}~\bibnamefont
  {Koumoutsakos}},\ }\href@noop {} {\bibfield  {journal} {\bibinfo  {journal}
  {Evolutionary computation}\ }\textbf {\bibinfo {volume} {11}},\ \bibinfo
  {pages} {1} (\bibinfo {year} {2003})}\BibitemShut {NoStop}%
\bibitem [{\citenamefont {B{\"u}cker}\ \emph {et~al.}(2013)\citenamefont
  {B{\"u}cker}, \citenamefont {Berrada}, \citenamefont {Van~Frank},
  \citenamefont {Schaff}, \citenamefont {Schumm}, \citenamefont {Schmiedmayer},
  \citenamefont {J{\"a}ger}, \citenamefont {Grond},\ and\ \citenamefont
  {Hohenester}}]{bucker2013vibrational}%
  \BibitemOpen
  \bibfield  {author} {\bibinfo {author} {\bibfnamefont {R.}~\bibnamefont
  {B{\"u}cker}}, \bibinfo {author} {\bibfnamefont {T.}~\bibnamefont {Berrada}},
  \bibinfo {author} {\bibfnamefont {S.}~\bibnamefont {Van~Frank}}, \bibinfo
  {author} {\bibfnamefont {J.-F.}\ \bibnamefont {Schaff}}, \bibinfo {author}
  {\bibfnamefont {T.}~\bibnamefont {Schumm}}, \bibinfo {author} {\bibfnamefont
  {J.}~\bibnamefont {Schmiedmayer}}, \bibinfo {author} {\bibfnamefont
  {G.}~\bibnamefont {J{\"a}ger}}, \bibinfo {author} {\bibfnamefont
  {J.}~\bibnamefont {Grond}}, \ and\ \bibinfo {author} {\bibfnamefont
  {U.}~\bibnamefont {Hohenester}},\ }\href@noop {} {\bibfield  {journal}
  {\bibinfo  {journal} {Journal of Physics B: Atomic, Molecular and Optical
  Physics}\ }\textbf {\bibinfo {volume} {46}},\ \bibinfo {pages} {104012}
  (\bibinfo {year} {2013})}\BibitemShut {NoStop}%
\bibitem [{\citenamefont {Schumm}\ \emph {et~al.}(2005)\citenamefont {Schumm},
  \citenamefont {Hofferberth}, \citenamefont {Andersson}, \citenamefont
  {Wildermuth}, \citenamefont {Groth}, \citenamefont {Bar-Joseph},
  \citenamefont {Schmiedmayer},\ and\ \citenamefont
  {Kr{\"u}ger}}]{schumm2005matter}%
  \BibitemOpen
  \bibfield  {author} {\bibinfo {author} {\bibfnamefont {T.}~\bibnamefont
  {Schumm}}, \bibinfo {author} {\bibfnamefont {S.}~\bibnamefont {Hofferberth}},
  \bibinfo {author} {\bibfnamefont {L.~M.}\ \bibnamefont {Andersson}}, \bibinfo
  {author} {\bibfnamefont {S.}~\bibnamefont {Wildermuth}}, \bibinfo {author}
  {\bibfnamefont {S.}~\bibnamefont {Groth}}, \bibinfo {author} {\bibfnamefont
  {I.}~\bibnamefont {Bar-Joseph}}, \bibinfo {author} {\bibfnamefont
  {J.}~\bibnamefont {Schmiedmayer}}, \ and\ \bibinfo {author} {\bibfnamefont
  {P.}~\bibnamefont {Kr{\"u}ger}},\ }\href@noop {} {\bibfield  {journal}
  {\bibinfo  {journal} {Nature physics}\ }\textbf {\bibinfo {volume} {1}},\
  \bibinfo {pages} {57} (\bibinfo {year} {2005})}\BibitemShut {NoStop}%
\bibitem [{\citenamefont {B{\"u}cker}\ \emph {et~al.}(2011)\citenamefont
  {B{\"u}cker}, \citenamefont {Grond}, \citenamefont {Manz}, \citenamefont
  {Berrada}, \citenamefont {Betz}, \citenamefont {Koller}, \citenamefont
  {Hohenester}, \citenamefont {Schumm}, \citenamefont {Perrin},\ and\
  \citenamefont {Schmiedmayer}}]{bucker2011twin}%
  \BibitemOpen
  \bibfield  {author} {\bibinfo {author} {\bibfnamefont {R.}~\bibnamefont
  {B{\"u}cker}}, \bibinfo {author} {\bibfnamefont {J.}~\bibnamefont {Grond}},
  \bibinfo {author} {\bibfnamefont {S.}~\bibnamefont {Manz}}, \bibinfo {author}
  {\bibfnamefont {T.}~\bibnamefont {Berrada}}, \bibinfo {author} {\bibfnamefont
  {T.}~\bibnamefont {Betz}}, \bibinfo {author} {\bibfnamefont {C.}~\bibnamefont
  {Koller}}, \bibinfo {author} {\bibfnamefont {U.}~\bibnamefont {Hohenester}},
  \bibinfo {author} {\bibfnamefont {T.}~\bibnamefont {Schumm}}, \bibinfo
  {author} {\bibfnamefont {A.}~\bibnamefont {Perrin}}, \ and\ \bibinfo {author}
  {\bibfnamefont {J.}~\bibnamefont {Schmiedmayer}},\ }\href@noop {} {\bibfield
  {journal} {\bibinfo  {journal} {Nature Physics}\ }\textbf {\bibinfo {volume}
  {7}},\ \bibinfo {pages} {608} (\bibinfo {year} {2011})}\BibitemShut {NoStop}%
\bibitem [{\citenamefont {Gerbier}(2004)}]{gerbier2004quasi}%
  \BibitemOpen
  \bibfield  {author} {\bibinfo {author} {\bibfnamefont {F.}~\bibnamefont
  {Gerbier}},\ }\href@noop {} {\bibfield  {journal} {\bibinfo  {journal} {EPL
  (Europhysics Letters)}\ }\textbf {\bibinfo {volume} {66}},\ \bibinfo {pages}
  {771} (\bibinfo {year} {2004})}\BibitemShut {NoStop}%
\bibitem [{\citenamefont {Caneva}\ \emph
  {et~al.}(2011{\natexlab{b}})\citenamefont {Caneva}, \citenamefont {Calarco},
  \citenamefont {Fazio}, \citenamefont {Santoro},\ and\ \citenamefont
  {Montangero}}]{caneva2011speeding}%
  \BibitemOpen
  \bibfield  {author} {\bibinfo {author} {\bibfnamefont {T.}~\bibnamefont
  {Caneva}}, \bibinfo {author} {\bibfnamefont {T.}~\bibnamefont {Calarco}},
  \bibinfo {author} {\bibfnamefont {R.}~\bibnamefont {Fazio}}, \bibinfo
  {author} {\bibfnamefont {G.~E.}\ \bibnamefont {Santoro}}, \ and\ \bibinfo
  {author} {\bibfnamefont {S.}~\bibnamefont {Montangero}},\ }\href@noop {}
  {\bibfield  {journal} {\bibinfo  {journal} {Physical Review A}\ }\textbf
  {\bibinfo {volume} {84}},\ \bibinfo {pages} {012312} (\bibinfo {year}
  {2011}{\natexlab{b}})}\BibitemShut {NoStop}%
\bibitem [{\citenamefont {Lesanovsky}\ \emph {et~al.}(2006)\citenamefont
  {Lesanovsky}, \citenamefont {Schumm}, \citenamefont {Hofferberth},
  \citenamefont {Andersson}, \citenamefont {Kr{\"u}ger},\ and\ \citenamefont
  {Schmiedmayer}}]{lesanovsky2006adiabatic}%
  \BibitemOpen
  \bibfield  {author} {\bibinfo {author} {\bibfnamefont {I.}~\bibnamefont
  {Lesanovsky}}, \bibinfo {author} {\bibfnamefont {T.}~\bibnamefont {Schumm}},
  \bibinfo {author} {\bibfnamefont {S.}~\bibnamefont {Hofferberth}}, \bibinfo
  {author} {\bibfnamefont {L.~M.}\ \bibnamefont {Andersson}}, \bibinfo {author}
  {\bibfnamefont {P.}~\bibnamefont {Kr{\"u}ger}}, \ and\ \bibinfo {author}
  {\bibfnamefont {J.}~\bibnamefont {Schmiedmayer}},\ }\href@noop {} {\bibfield
  {journal} {\bibinfo  {journal} {Physical Review A}\ }\textbf {\bibinfo
  {volume} {73}},\ \bibinfo {pages} {033619} (\bibinfo {year}
  {2006})}\BibitemShut {NoStop}%
\end{thebibliography}%

\end{document}